\documentclass[12pt]{iopart}

\usepackage{color}
\usepackage{bm}
\usepackage{citesort}
\usepackage{graphicx}
\usepackage{amssymb}
\usepackage[dvipsnames]{xcolor}

\def\be{\begin{equation}}
\def\ee{\end{equation}}
\def\bea{\begin{eqnarray}}
\def\eea{\end{eqnarray}}

\def\ptl{\partial}

\usepackage{citesort}

\begin{document}
\nocite{*}

\title{Cosmological
singularity as an informational seed for Everything}

\noindent \qquad June 21, 2021

\author{S.L. Cherkas\dag   \ and V.L. Kalashnikov\ddag}

\address{\dag\
Institute for Nuclear Problems, Bobruiskaya 11, Minsk 220030,
Belarus}

\address{\ddag\ Facolt\'a
di Ingegneria dell'Informazione, Informatica e Statistica,
Sapienza Universit\'a di Roma, Via Eudossiana 18 00189 - Roma, RM,
Italia}

\begin{abstract} It is shown how to place some amount of matter into the
cosmological singularity and to encode its state. A free and
massless scalar field is considered as a prototype of matter. Two
different but coherent approaches to this issue are presented. The
expression for the scalar particles' spectral energy density,
which is initially encoded at the singularity, is deduced. An
informational aspect of the problem is discussed.
\end{abstract}

\section{Introduction}

The well-known Penrose theorem
\cite{Penrouse1965,Geroch1968,Hawking1970} states that, under
quite general conditions, the initial point of the universe
evolution should be singular. Removing of a singularity on a
classical level requires some ``bounces"
\cite{Minkevich2006,Brand2017}, using exotic matter or modified
theories of gravity. Such ``bounces'' take place also in some
quantum gravity theories, for instance, in the loop quantum
gravity \cite{Ashtekar2006,Singh2009}. Another view on a
cosmological singularity could be that it is not a problem but a
necessary feature, related rather with the concept of time
\cite{hans,ill} than with the gravity. Although one could imagine
some infinite events in time, everything that we could observe
experimentally has the beginning and the end. In such a case, the
beginning of the universe could be associated with a cosmological
singularity.

The universe singularity could remain in the quantum cosmology, as
well \cite{Cherkas2006,Cherkas2012,Cherkas2015}. Moreover, it is
appealing to set the initial conditions for the universe evolution
at the singularity per se \cite{Cherkas2017}.

%%%%%%%%%%%%%%% End of first page %%%%%%%%%%%%%%%%%%%%%

\maketitle

This issue includes the origin of matter and information in the
universe. The mainstream view suggests inflation
\cite{Starobinsky1980,Guth1981,Liddle} as the ingredient of the
cosmological theory, despite the intensive discussions regarding
the inflation (see the so-called ``letter of 33th'' \cite{open}).
At the end of inflation, the matter appears from the inflanton
field decay, which occurs when it began to oscillate
\cite{Linde1990,Mukhanov2005}. It is not an invulnerable point of
view because the inflaton field cannot be associated with the
known fields of the Standard Model of particle physics. Also in
the context of the information, inflation erases all the previous
information, up to the scales of so-called trans-Planckian physics
\cite{trans1,trans2,trans3} (see, e.g., \cite{Agullo2018}), and
generates a spectrum of the initial inhomogeneities from vacuum
fluctuations of the inflaton field and metric tensor. For the
theories without inflation, another explanation for the matter
origin in the universe is needed.

Since the most theories without inflation predict ``bounces,''
i.e., exclude singularity \cite{bran}, the content of the present
paper implies a relatively narrow class of the theories with the
singularity. Still, without inflation, otherwise, there would be
no sense to talk about information if the inflation rewrites it.
The Milne-like universes could be an example of the model
\cite{mil,John,loh1,conf0,levy,fm,lew,Singh,plasm}, where a stage
of the linear universe expansion in cosmic time solves the problem
of the horizon without inflation.

As was shown earlier \cite{Cherkas2017}, some finite quantities
exist at the singularity, namely, the momentums of the dynamical
variables despite the dynamical variables' infiniteness. For
instance, the scalar fields' amplitudes are infinite at the
singularity, but their momentums are finite. In the quantum
picture, momentums' finiteness allows building a wave packet at
the singularity and setting an initial condition for the universe
evolution. It was shown at the example of the Gowdy model
considered in the quasi-Heisenberg picture \cite{Cherkas2017}.
Here, in the first section, we demonstrate it using the familiar
approach to the quantum fields on the classical background. In the
second section, as a bridge to the complete quantization of
gravity, we consider a toy model where the scale factor evolution
is uncoupled from the universe stuffing. It is shown that both
approaches give the same results, i.e., reflect the same physical
reality. Finally, we discuss the information context of the
initial universe state.

\section{Quantum fields on the classical background}

The main idea that the field momentums remain finite at the
singularity and this allows setting the momentum wave packet
defining the evolution of the system could be realized by the the
conventional approach to the quantum fields on the classical
background \cite{Birrell1982}. The interval for isotropic, a
uniform and the flat universe is

\be
ds^2\equiv g_{\mu\nu}dx^\mu dx^\nu
=a^2(\eta)\left(d\eta^2-\tilde\gamma_{ij}dx^idx^j\right),
\label{int}
\ee
where $a$ is a scale factor, $\eta$ is a conformal time and $
\tilde\gamma _{ij}=\mbox{diag}\{1,1,1\}$ is an Euclidean 3-metric.
A scalar field $\phi(\eta,\bm r)$ on this background can be
represented in the form of the Fourier series $\phi(\eta,\bm
r)=\sum_{\bm k} \varphi_{\bm k}(\eta) e^{i {\bm k}\bm r}$.

The Hamiltonian describing an evolution of the $\bm k-$modes
\cite{Birrell1982} is written as
\be
H=\sum_{\bm k}\frac{\pi_{\bm k}\pi_{ -\bm
k}}{2a^{2}}+\frac{1}{2}a^{2}k^2\varphi_{\bm k}\varphi_{-\bm k}.~~~
\label{hsm0}
\ee
The creation and annihilation operators can be introduced as
\begin{equation}
\hat \varphi_{\bm k}=\hat {\mbox{a}}^+_{-\bm k}u_{k}^*(\eta)+\hat
{\mbox{a}}_{\bm k} u_{k}(\eta). \label{ff}
\end{equation}
The functions  $u_{k}$ satisfy
\bea
u^{\prime\prime}_{k}+\frac{2 a^\prime}{a}u^\prime_{k} +k^2 u_{
k}=0,
\label{vf}\\
a^2(\eta)(u_k(\eta)
\,{u_k^\prime}^*(\eta)-u_k^*(\eta)\,u_k^\prime(\eta))=i,
\label{rel2}
\eea
where the last equation (\ref{rel2}) is a consequence of the
canonical commutators $[\hat \pi_{\bm k},\hat \varphi_{\bm
q}]=-i\delta_{\bm k,\bm q}$ for the field  (\ref{ff}) and the
momentum
\be
\hat \pi_{\bm k}=a^2\hat \varphi^{\prime}_{-\bm k}=a^2(\eta)(\hat
{\mbox{a}}_{-\bm k}u_{k}^{\prime}(\eta)+\hat {\mbox{a}_{\bm k}^+}
u_{k}^{*\prime}(\eta)).
\label{mom}
\ee

 Let us show that the momentum (\ref{mom}) is finite at  $\eta=0$. In the
vicinity of singularity, the function $u_k (\eta )$ satisfies Eq.
(\ref{vf}) asymptotically without the last term. This equation can
be converted into the form

\begin{equation}
 \frac{d}{d\eta}\left( a(\eta )^2u^\prime_k(\eta) \right) = 0.
\label{vf1}
\end{equation}
From Eqs. (\ref{mom}) and (\ref{vf1}) one may conclude that the
momentums $\hat {\pi }_{\rm {\bf k}} $ are asymptotically some
constant operators in the vicinity of singularity. The assumption,
that the kinetic terms $u^{\prime\prime}_{k}+\frac{2
a^\prime}{a}u^\prime_{k}$ are dominant over $k^2 u_{ k}$ in the
vicinity of singularity, is valid, e.g., for the dependencies
$a(\eta ) \sim \eta ^n$, when $u_k\approx \eta^{1-2n}$. In
particular, these dependencies include $a(\eta ) \sim \eta $
(radiation background) and $a(\eta ) \sim \eta ^2$ (matter
background). The creation $\hat {\mbox{a}}_{\bm k}^+$ and
annihilation $\hat {\mbox{a}}_{\bm k}$ operators are very
convenient instruments for the description of the quantum fields
at the late times, when the field oscillators oscillate and the
term $k^2{\hat \varphi}_{\bm k}$ exceeds $\frac{a^\prime}{a}{\hat
\varphi}_{\bm k}^\prime$ one. However, at early times, it is more
convenient to describe fields in terms of the eigen functions of
the operators $\hat \pi_k(0)=\hat P_{\bm k}$. Let us define the
time-independent operator
\be
\hat P_{\bm k}=\alpha_{k}\hat {\mbox{a}}_{-\bm k}+\alpha_{k}^*\hat
{\mbox{a}_{\bm k}^+},
\label{p}
\ee
where the complex-valued constants are
\be
\alpha_k=a^2(\eta) u_k^\prime(\eta)\bigr|_{\,\eta\rightarrow 0},
\label{con}
\ee
and let's introduce an additional constant operator
\be
\hat X_{\bm k}=b_{k}\left(\hat {\mbox{a}}_{\bm k}+\hat
{\mbox{a}_{-\bm k}^+}\right),
\label{x}
\ee
where $b_k$ is some real constant, which must be defined. The
requirements of the fulfilment of the commutation relations
 \be
[\hat P_{\bm k},\hat X_{\bm q}]=-i\delta_{\bm k,\bm q},
 \ee
with the taking into account of $[{\hat{\mbox{a}}}_{\bm k},\,
{\hat{\mbox{a}}}^+_{\bm k}]=1$ allows obtaining the real constants
$b_k$
\be
b_{k}=-\frac{i}{\alpha_k  -\alpha^*_k }.
\ee

The phases of the quantities $\alpha_k$ are determined by the
phases of the functions $u_k(\eta)$. These phases are not formal
parameters but influence a quantum state of a system \cite{proc}.
Then, the creation and annihilation operators can be expressed
with the help of  Eqs. (\ref{p}), (\ref{x}) through $\hat X_{\bm
k}$ and $\hat P_{\bm k}$\ as
\bea
{\hat{\mbox{a}}}_{\bm k}=\frac{\hat P^+_{\bm k}}{\alpha_k
-\alpha_k^*
 }-i \alpha^*_k \hat X_{\bm k},\nonumber\\
{\hat{\mbox{a}}^+}_{-\bm k} =\frac{\hat P^+_{\bm k}}{ \alpha_k^*
-\alpha_k}+i \alpha_k \hat X_{\bm k}.
\label{aa}
\eea
Substitution  (\ref{aa}) into (\ref{ff}) allows writing the field
operator in the terms of operators $\hat P_{\bm k}$, $\hat X_{\bm
k}$:
\bea
\hat \varphi_{\bm k}(\eta)=\frac{\hat P_{\bm k}^+}{\alpha_k
-\alpha^*_k} \left(u_k(\eta )- u^*_k(\eta )\right)+i\hat X_{\bm
k}(\alpha_k u^*_k(\eta )-\alpha^*_k u_k(\eta )).~
\label{pole}
   \eea

The momentum operator $\hat \pi_{\bm k}(\eta)$ can be expressed
analogously
\bea
\fl \hat\pi_{\bm k}(\eta)=a(\eta)^2\biggl(\frac{\hat P_{\bm
k}}{\alpha_k-\alpha_k^*}\left(u_k^\prime(\eta)-u_k^{*\prime}(\eta)\right)+i
\hat X^+_{\bm k}\left(\alpha_k
u_k^{*\prime}(\eta)-\alpha_k^*u_k^{\prime}(\eta)\right)\biggr),~
\eea
where it is taken into account that $\hat X_{-\bm k}=\hat X^+_{\bm
k}$, $\hat P_{-\bm k}=\hat P^+_{\bm k}$. From
(\ref{rel2}),(\ref{con}),(\ref{pole}) it follows that $\hat X_{\bm
k}$ is a nonsingular part of $\hat \varphi_{\bm k}(\eta)$ at
$\eta=0$.

The mean value of an arbitrary operator over a wave packet is
written as
\bea
\fl <C|\hat A\left(\eta,\left\{P_{\bm k},i\frac{\ptl}{\ptl P_{\bm
k}},P_{\bm k}^*,i\frac{\ptl}{\ptl P_{\bm
k}^*}\right\}\right)|C>=~~~~~~~~~~~~~~~~~~~~\nonumber\\\int
(C(\{P_{\bm k},P_{\bm k}^*\}))^*\hat A \,C(\{P_{\bm k},P_{\bm
k}^*\})\mathcal D P_{\bm k}\mathcal D P_{\bm k}^*,~
\label{arb}
\eea
where  $\mathcal D P_{\bm q}\equiv dP_{\bm k_1}dP_{\bm k_2}...$
 and the following realization of the operators is implied:
\be
\hat P_{\bm k}=P_{\bm k},~~~ \hat P_{\bm k}^+=P_{\bm k}^*,~~~\hat
X_{\bm k}=i\frac{\ptl }{\ptl P_{\bm k}},~~~\hat X_{\bm
k}^+=i\frac{\ptl }{\ptl P_{\bm k}^*}.
\ee

 In particular, the
mean value of $\hat \varphi_{\bm k}(\eta)$ over a wave packet
looks as
\bea
\fl <C|\hat \varphi_{\bm k}(\eta)|C>=\frac{u_k(\eta )- u_k^*(\eta
)}{\left(\alpha_k- \alpha^*_k\right)}\int (C(\{P_{\bm q},P_{\bm
k}^*\}))^*P^*_{\bm k}C(\{P_{\bm q},P_{\bm k}^*\})\mathcal D P_{\bm
q}\mathcal D P_{\bm
q}^*-\nonumber\\
\left(\alpha_k u_k^*(\eta)-u_k(\eta) \alpha^*_k\right)\int
(C(\{P_{\bm q},P_{\bm k}^*\}))^*\frac{\ptl}{\ptl P_{\bm
k}}C(\{P_{\bm q},P_{\bm k}^*\})\mathcal D P_{\bm q}\mathcal D
P_{\bm q}^*.~~~~
\label{22}
\eea
Let one has some fixed set of the functions $u_k$ (in the paper it
was used a set in which quantities $\alpha_k$ is pure imaginary)
and intends to proceed to another set $u_k e^{-i\theta k}$. Then
formula (\ref{pole})  will look as
\be
\hat \varphi_{\bm k}(\eta)=\frac{\hat P_{\bm k}^+ \left(e^{- i
\theta_k }u_k(\eta )-e^{ i \theta_k } u^*_k(\eta )\right)}{e^{- i
\theta_k }\alpha_k -\alpha^*_k e^{ i \theta_k }} +i \hat X_{\bm k}
(\alpha_k u^*_k(\eta )-\alpha^*_k u_k(\eta )),
\label{pole1}
   \ee
where
  $\hat
P_{\bm k}=P_{\bm k}$ and $\hat X_{\bm k}=i\frac{\ptl}{\ptl P_{\bm
k}}$. The  mean value of  $\hat \varphi_{\bm k}(\eta)$ over wave
packet takes the form
\bea
\fl <C|\hat \varphi_{\bm k}(\eta)|C>=\frac{e^{- i \theta_{k}}
u_k(\eta )-e^{ i \theta_{k} } u_k^*(\eta )}{e^{- i
\theta_{k}}\alpha_k-e^{ i \theta_{ k} } \alpha^*_k}\int (C(P_{\bm
q}))^*P^*_{\bm k}C(P_{\bm q})\mathcal D P_{\bm q}\mathcal D P_{\bm
q}^*\nonumber\\
- \left(\alpha_k u_k^*(\eta)-u_k(\eta) \alpha^*_k\right)\int
(C(P_{\bm q}))^*\frac{\ptl}{\ptl P_{\bm k}}C(P_{\bm q})\mathcal D
P_{\bm q}\mathcal D P_{\bm q}^*,
\label{222}
\eea
instead of the expression (\ref{22}).

 Let us
consider the transformation of the wave packet $ C(P_{\bm
k})\rightarrow C(P_{\bm k})\exp\left(-i\sum_{\bm q} g_{ q}P^*_{\bm
q}P_{\bm q}\right)$ in the Eq. (\ref{22}), where $g_{k}$ is some
real constant.  It could be seen from (\ref{22}),(\ref{222}) that
the transformation above is equivalent to the appearence the phase
$\theta_{k}$ satisfying to the equation
\be
\frac{e^{i\theta_k}}{e^{-i\theta_k}\alpha_k-e^{i\theta_k}\alpha_k^*}=\frac{1}{\alpha_k-\alpha_k^*}-i
g_k\alpha_k^*.
\label{3}
   \ee
That is, the phases $\theta_k$ have to be considered twofold:
mathematically, they are the phases of the basis functions
$u_k(\eta)$, but physically,  they are  the property of the
quantum state, because  they are equivalent  to the  constants
$g_k$.

\section{Quasi-Heisenberg picture for matter appearance}

As a gravity quantization is a too complicated issue, let us
consider a toy quantum model based on an analogy with the Gowdy
cosmological model \cite{Cherkas2017}, where the Hamiltonian
contains the gravitational degrees of freedom in the first order
on momentums. To exclude the quantum field back-reaction to the
universe expansion, the heuristic Hamiltonian could be written as
\be
H=-p_a\,f(a)+ \sum_{\bm k}\frac{\pi_{\bm k}\pi_{ -\bm
k}}{2a^{2}}+\frac{1}{2}a^{2}k^2\varphi_{\bm k}\varphi_{-\bm k},~~~
\label{hsm}
\ee
where $f(a)$ is some arbitrary function of the scale factor,
$\pi_{\bm k}$ is the momentums corresponding to $\varphi_{\bm k}$.
Let us consider Eq. (\ref{hsm}) not only as a Hamiltonian but
simultaneously as a constraint $H=0$, again, by analogy with the
Gowdy model.

As one can see, the Hamiltonian (\ref{hsm}) contains momentum
$p_a$ corresponding to the universe scale factor $a$, as it occurs
in the typical minisuperspace models. However, it has a
first-order degree like the Gowdy model \cite{Cherkas2017}. Note
that the Hamiltonian (\ref{hsm}) is purely heuristic and differs
from that coming from the scalar constraint of GR, which is
second-order in momentums.

The Hamilton equations result in the equation of motion for the
scale factor
\be
a^\prime=-\frac{\ptl H}{\ptl p_a}=f(a).
\label{am}
\ee

The scale factor is classical and given by the solution of Eq.
(\ref{am})
\be
\int_{a_{0}}^{a(\eta)}\frac{d\varsigma}{f(\varsigma)}=\eta,
\label{a0}
\ee
so that  $a(0)=a_{0}$. For instance, $f(a)=const=\mathcal H$ gives
\be
a(\eta)=\mathcal H \eta+a_{0},
\label{exmp}
\ee
where $a_{0}$ is the initial value of the scale factor at
$\eta=0$. Namely, this particular case corresponding to the
radiation domination universe will be considered below for
illustration.

Let us sketch the quantization scheme in the quasi-Heisenberg
picture described in
\cite{Cherkas2006,Cherkas2012,Cherkas2015,Cherkas2017}. It
consists of the quantization of the classical equations of motion,
i.e., one should write ``hats'' under every quantity in the
equations of motion. Then, one has to define the commutation
relations for the operators, to chose the operator ordering in the
equations, and finally to define the Hilbert space, where the
operators act.

The equations of motion for the scalar field is written as

\be
\varphi_{\bm k}^\prime=\frac{\ptl H}{\ptl \pi_{\bm
k}}=\frac{\pi_{-\bm k}}{a^2},~~~~\pi_{\bm k}^\prime=-\frac{\ptl
H}{\ptl \varphi_{\bm k}}=-a^2k^2\varphi_{-\bm k}. \label{heq}
\ee
The resulting equation of motion originates form (\ref{heq}):
\be
\varphi^{\prime\prime}_{\bm k}+\frac{2
a^\prime}{a}\varphi^\prime_{\bm k} +k^2 \varphi_{\bm k}=0.
\label{vf2}
\ee

The solution of Eq. (\ref{vf2}) is convenient to write through the
functions  $u_k$ satisfying Eqs. (\ref{vf}) and (\ref{rel2}):
\be
\fl \hat \varphi_{\bm k}(\eta)=i\left(u_k(\eta ) \left(\hat
P^+_{\bm k} u^*_k(0)-a_0^2 \hat \Phi_{\bm k}
u^{*\prime}_k(0)\right)-u^*_k(\eta) \left(\hat P^{+}_{\bm k}
u_k(0)-a_0^2 \hat \Phi_{\bm k} u_k^\prime(0)\right)\right),
\label{meth1}
\ee
where the operators $\hat P_{\bm k}$ and $\hat \Phi_{\bm k}$ do
not depend on time and satisfy  the commutation relations $[\hat
P_{\bm k},\hat \Phi_{\bm q}]=-i\delta_{\bm k,\bm q}$. They are the
initial values of the operators $\hat \pi_{\bm k}(\eta)$ and $\hat
\varphi_{\bm k}(\eta)$ at $\eta=0$. In the momentum representation
$\hat P_{\bm k}=P_{\bm k}$, $\hat \Phi_{\bm k}=i\frac{\ptl}{\ptl
P_{\bm k}}$. It is not difficult to see that $\hat \pi_{\bm
k}(\eta)$ and $\hat \varphi_{\bm k}(\eta)$ have the commutation
relation $[\hat \pi_{\bm k},\hat \phi_{\bm q}]=-i\delta_{\bm k,\bm
q}$ due to Eq. (\ref{rel2}).

Momentum $p_a$ is expressed from the constraint $H=0$ as
\be
\hat p_a=\frac{1}{f(a)}\left(\sum_{\bm k}\frac{\hat \pi_{\bm
k}\hat\pi^+_{ \bm k}}{2a^{2}}+\frac{1}{2}a^{2}k^2\hat \varphi_{\bm
k}\hat \varphi^+_{\bm k}\right),
\ee
where it is taken into account that $\hat \varphi_{-\bm k}=\hat
\varphi^+_{\bm k}$.

According to the quasi-Heisenberg scheme, the next step is to
build the Hilbert space where the quasi-Heisenberg operators act.
That can be done with the help of the Wheeler-DeWitt equation in
the vicinity of the small scale factors $a\sim a_0\rightarrow 0$.
At this stage anew, we quantize the Hamiltonian constraint
canonically
 \cite{Cherkas2006,Cherkas2012,Cherkas2015,Cherkas2017}, which gives in the momentum representation
\be
i f(a) \ptl_a\psi(a,\{P_{\bm k},P^*_{\bm
k}\})=\left(\frac{1}{a^2}\sum_{\bm q, {q_z>0}}P_{\bm q}P_{\bm q}^*
\right)\psi (a,\{P_{\bm k},P^*_{\bm k}\}),
\label{dev}
\ee
where only half-space of the wave vectors $\bm q$ is taken by
condition $q_z>0$  to avoid a double counting originating from
$P_{-\bm k}=P^*_{\bm k}$. The curly bracket  in (\ref{dev})
denotes a set of $P_{\bm k_1},P^*_{\bm k_1},P_{\bm k_2},P^*_{\bm
k_2}...$. The solution of Eq. (\ref{dev}) is

\be
\fl \psi (a,\{P_{\bm k},P^*_{\bm k}\})=\exp\left(-i\sum_{\bm q,
q_z>0}P_{\bm q}P_{\bm q}^*\left(\int\frac{da}{a^2f(a)}+\Theta_{\bm
q}\right)\right)C(\{P_{\bm k},P^*_{\bm k}\}),
\label{psi}
\ee
where $\Theta_{\bm k}$ is an arbitrary real constant arising as a
result of integration and $C(\{P_{\bm k},P^*_{\bm k}\})$ is the
momentum wave packet defining the quantum state of the model and
containing all the information about it.

The calculation of the mean values \cite{Cherkas2017} includes the
integration over $\mathcal D P_{\bm k}$, $\mathcal D P^*_{\bm k}$
on the hypersurface $a=a_0$ and proceeding the limit
$a_0\rightarrow 0$:
\bea
\fl <\psi|\hat A\left(\eta,a,\left\{P_{\bm k},i\frac{\ptl}{\ptl
P_{\bm k}},P_{\bm k}^*,i\frac{\ptl}{\ptl P_{\bm
k}^*}\right\}\right)|\psi>=~~~~~~~~~~~~~~~~~\nonumber\\\int
\psi^*(a,\{P_{\bm k},P_{\bm k}^*\})\hat A \,\psi(a,\{P_{\bm
k},P_{\bm k}^*\})\mathcal D P_{\bm k}\mathcal D P_{\bm
k}^*\biggr|_{\,a=a_0\rightarrow 0}.~
\label{me}
\eea

 In particular, the mean value calculation $\hat
\varphi_{\bm k}(\eta)$ leads to
\bea
\fl <\psi|\hat \varphi_{\bm k}(\eta)|\psi>= i \bigl(a_0^2
\left(I(a_0) \, +\Theta_{\bm k}\right) \left(u^\prime_k(0)
u^*_k(\eta )-u_k(\eta )
u_k^{*\prime}(0)\right)~~~~~~~~~~\nonumber\\+u_k^*(0) u_k(\eta
)-u_k(0) u_k^*(\eta )\bigr)\int (C(\{P\})\}))^*P^*_{\bm
k}C(\{P\})\mathcal D P_{\bm q}\mathcal D P_{\bm q}^*~~~~~~~~~~~\nonumber\\
-a_0^2 \bigl(u_k^\prime(0) u_k^*(\eta)-u_k(\eta)
u_k^{*\prime}(0)\bigr)\int (C(\{P\})^*\frac{\ptl}{\ptl P_{\bm
k}}C(\{P\})\mathcal D P_{\bm q}\mathcal D P_{\bm
q}^*\biggr|_{\,a=a_0\rightarrow 0},\nonumber\\
\label{11}
\eea
where the indefinite integral is defined as $I(a)=\int
\frac{1}{a^2 f(a)}da$ and $C(\{P_{\bm q},P_{\bm q}^*\})$ is
implied under C(\{P\}).

It should be noted that, generally, the quasi-Heisenberg
quantization scheme holds also in a
 more general case, when the background geometry is quantum \cite{Cherkas2006,Cherkas2012,Cherkas2015} and
 the Klein-Gordon ``current'' scalar product appears instead of the Schr\"{o}dinger-like product in (\ref{me}). However,
 Eq. (\ref{me}) is not a pure Schr\"{o}dinger ``density'' scalar product (i.e., more exactly, an expression
 for the mean value calculation)  even in the present case, but also it includes hyperplane $a=a_0\rightarrow 0$ explicitly.

 As may see, the resulting Eq. (\ref{11}) is analogous to Eq. (\ref{22}). Considering the
functions $u_k(\eta)$ in (\ref{22}) as a limit $a_0\rightarrow 0$
of the functions $u(\eta,a_0)$ in (\ref{11}) allows concluding
that these formulas coincide exactly if the constants $\Theta_k$
are defined as
 \be
\Theta_{k}=\frac{i\left(1+2u^*_k(0,a_0)|\alpha_k|\right)}{2|\alpha_k|^2}-I(a_0)\biggr|_{a_0\rightarrow
0}\label{thet}~~,
 \ee
where we conventionally define the functions $u_k(\eta,a_0)$ in
such a way  that $\alpha_k/|\alpha_k|=i$.

Let's come to a concrete example of (\ref{exmp}), where
$I(a_0)=-\frac{1}{\mathcal H a_0}$ and the functions are
\bea
u_k(\eta,a_0)=-\frac{ i\sqrt{a_0^2 k^2+\mathcal H^2}}{\sqrt{2k}
(a_0+\eta
\mathcal H) (\mathcal H+i a_0 k)}e^{-ik \eta},\nonumber\\
u_k^*(\eta,a_0)=\frac{ i\sqrt{a_0^2 k^2+\mathcal H^2}}{\sqrt{2k}
(a_0+\eta \mathcal H) (\mathcal H-i a_0 k)}e^{ik \eta},
\label{uuu}
\eea
that are in the limit of $a_0\rightarrow 0$
\bea
u_k(\eta)=-\frac{ i}{\sqrt{2k} \,\eta
\mathcal H }e^{-ik \eta},\nonumber\\
u_k^*(\eta)=\frac{ i}{\sqrt{2k} \,\eta \mathcal H }e^{ik \eta},
\label{uu}
\eea
and $\alpha_k=i\mathcal H^2/\sqrt{2k}$.

The evaluation of (\ref{thet}) gives $\Theta_k=0$. Further
calculations show that two formalisms
 are fully equivalent not only for the mean value of $\hat \varphi_{\bm k}(\eta)$, but for
  other operators mean values, as well.

\section{Energy density spectrum of the created particles}

Although the field oscillators do not oscillate near $\eta\sim 0$,
the information about a matter is encoded at the singularity,
whereas the particle density becomes well-defined later when the
notion of a ``particle'' takes shape.

Let us consider a Gaussian wave packet
\be
C(\{P\})=N\prod_{\bm q}\exp\left(-\Delta_{\bm q}P_{\bm q}P_{\bm
q}^*\right),
\label{pak}
\ee
where $N$ is a normalizing constant and the function $\Delta_{\bm
q}$ has both real and complex parts $\Delta_{\bm q}=\Delta_{\bm
q}^\prime+i\Delta_{\bm q}^{\prime\prime}$. The quantities
$\Delta_{\bm q}^\prime$ should be positive to provide the integral
convergence.

The mean energy density of the created scalar particles can be
defined as

\be
\bar{\rho}=<C(\{P\})|\hat \rho |C(\{P\})>-<0|\hat \rho |0>,
\label{mat}
\ee
where
\be
\fl \hat \rho=\frac{1}{V}\int_V\left(\frac{\hat \phi^{\prime 2}}{2
a^2}+\frac{(\bm \nabla \hat \phi)^2}{2 a^2}\right)d^3\bm
r=\frac{1}{2a^2}\sum_{\bm k}{ \hat \varphi_{\bm k}^\prime\hat
\varphi_{ -\bm k }^\prime}+ k^2\hat \varphi_{ {\bm k}}\hat
\varphi_{- {\bm k}}\equiv\sum_{\bm k}\hat\wp_{\bm k}
\label{36}
\ee
and the vacuum energy density is
\be
\tilde \rho\equiv <0|\hat \rho
|0>=\frac{1}{a(\eta)^4}\left(\sum_{\bm
k}\frac{k}{2}+\frac{1}{4k\eta^2}\right)\equiv \sum_{\bm k} \tilde
\wp_{k}.
\label{vac}
\ee

The definition (\ref{mat}) includes both some state $|C>$ set at
singularity and a late time vacuum state $|0>$. The last one is an
ordinary vacuum state $a_{\bm k}|0>=0$, but, as shown later, this
state corresponds to some particular wave packet defined at the
singularity that does not produce particles in the future. It
could be said that everything is planted at the singularity.
However, to understand what will grow from this state, one needs
to have a late times ``measure,'' i.e., a well-defined vacuum.
Here, a purely classical background is considered, which allows
finding an analog of the vacuum state in the form of the wave
packet defined at the singularity. In the case of the common
quantum gravity, it is possible to define only the approximate
vacuum state at the late time when the background becomes
approximately classical.

According to Eq. (\ref{36}), the mean energy density consists of
the sum over every wave mode, and the quantity $\wp_k$ can be
considered as the spectral energy density of the created
particles. By turning from summation to integration, one comes to

\be
\bar \rho\equiv\sum_{\bm k}\bar\wp_{k}=\frac{1}{2\pi^2} \int
\bar\wp_{k}k^2dk,
\label{def}
\ee
 if $\bar \wp_{k}$ does not depend on
the direction of $\bm k$.

\begin{figure}[th]
\includegraphics[width=12.7 cm]{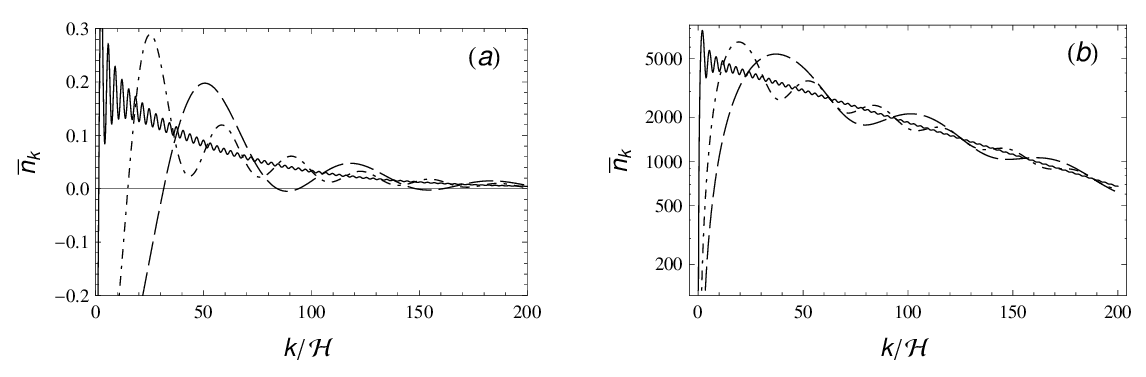}
\caption{\normalfont Spectral density of the particles $\bar
n_k=\bar\wp_k/k$  appeared  at the different moments of the
conformal time for some model functions $\Delta_{\bm
k}^{\prime\prime}$ and $\Delta_{\bm k}^\prime$  given by
(\ref{spec}). (a) $L=10^{-2}/\mathcal H$, $\Omega=1$,(b)
$L=10^{-2}/\mathcal H$, $\Omega=10^4$, where $\Omega$ and $L$ are
parameters describing wave packet (\ref{spec}).
 Dashed, dashed-dotted lines and
solid line corresponds to $\eta\mathcal H=0.05$, $\eta\mathcal
H=0.1$ and $\eta\mathcal H=1$, respectively. }\label{fig1}
\end{figure}

Let us expose the results on the mean value calculation of the
energy density (\ref{36}) for the wave packet (\ref{pak}) and the
dependence $a(\eta)=\mathcal H \eta$. The method of the second
section suggests using Eqs.(\ref{pole}), (\ref{arb}), (\ref{uu}).
Since the wave packet (\ref{36}) represents the
 product over $\bm k$-terms, the calculation of the energy density, which is a sum over the modes,
can be done for each $\bm k$-mode separately. The integration over
$dP_{\bm k}dP_{\bm k}^*$ is understood like that in the
holomorphic representation \cite{Faddeev1987}: $dP_{\bm k}dP_{\bm
k}^*\equiv{\rho_{\bm k} d\rho_{\bm k} d\theta_{\bm k}}$, where
$P_{\bm k}=\rho_{\bm k} e^{i\theta_{\bm k}}$.

The method of the  third section includes Eqs. (\ref{exmp}),
(\ref{meth1}), (\ref{psi}), (\ref{me}), (\ref{uuu}) and the limit
of $a_0\rightarrow 0$. It turns out to be that the different
terms, which diverge at $a_0\rightarrow 0$, cancel each other in
this limit. This topic was discussed earlier in
\cite{Cherkas2012}.

Both methods give the same result:
\bea
\fl a^4\bar \rho=\frac{1}{8}   \eta ^{-2}\mathcal H^{-2}\sum_{\bm
k}k^{-2}\Delta^{\prime-1}_{\bm k} \biggr(\left(2 \eta ^2
k^2+1\right) \left(\Delta^{\prime\prime 2}_{\bm k} \mathcal
H^4+\left(k-\Delta^\prime_{\bm k}
  \mathcal H^2\right)^2\right)+~~~~~~~~~~~~~~~~~~~~~~~~~~~~~
  \nonumber\\
  \fl 2 k \sin (2 \eta  k)
  \left(\eta  k^2-\mathcal H^2 \left(\eta
\mathcal H^2 |\Delta_{\bm k}|^2+\Delta^{\prime\prime}_{\bm k}
\right)\right)+
  \cos (2 \eta  k) \left(k^2 \left(4 \eta
\Delta^{\prime\prime}_{\bm k} \mathcal H^2+1\right)-\mathcal H^4
|\Delta_{\bm k}|^2\right)\biggr).~
\label{all}
\eea

At the late times $\eta\rightarrow \infty$, the oscillations decay
and
\bea
a^4\bar \rho\equiv a^4\sum_{\bm k}\bar \wp_k \approx \sum_{\bm
k}\frac{ \mathcal H^4|\Delta_{\bm k}|^2-2\mathcal H^2k\Delta_{\bm
k}^\prime+k^2}{4 \Delta^\prime_{\bm k} \mathcal H^2},
\label{stat}
\eea
which turns to zero at $\Delta_{\bm k}^\prime=k/\mathcal H^2$,
$\Delta_{\bm k}^{\prime\prime}=0$ in the agreement with
(\ref{all}). Thus, the particles do not appear this value of
$\Delta_{\bm k}$, and, thereby, it
  corresponds to a vacuum
state.

Let us consider some numerical example and take
\be
\Delta_{\bm k}^{\prime}=\frac{\sqrt{k^2+\mathcal H^4\Delta_{\bm
k}^{\prime\prime2}}}{\mathcal H^2},~~~\Delta_{\bm
k}^{\prime\prime}=\frac{\Omega k}{\mathcal H^2}\exp\left(-k
L\right),
\label{spec}
\ee
where $\Omega$ and $L$ are some constants. Parameter $\Omega$
increases oscillatory component of the wave packet, which leads to
increasing of the number of the created particles. The spectral
density of the particle number for this illustrative example is
shown in Fig. \ref{fig1}. The number of the created particles
increases with the increasing of $\Omega$ as it is shown in Fig.
\ref{fig1}.

The spectral density, as can be seen, is not always positive at
the early time, because the notion of ``particle'' is not
well-defined yet, but this quantity becomes positive later.

The mathematical aspect of the calculations requires some
additional comments. Integration $\mathcal D P_{\bm q}\mathcal D
P^*_{\bm q}\equiv dP_{\bm k_1} dP^*_{\bm k_1}dP_{\bm k_2}dP^*_{\bm
k_2}...$
 represents, in fact, a continual integration without a rigorously
defined measure\footnote{Nevertheless, that does not prevent to
deal formally with the Gaussian continual integrals
 \cite{Faddeev1987}. }. There are two possibilities to proceed with these integrals. The first possibility is to consider these integrals as finite-dimensional formally,  i.e., look at $|\bm k|$  as a quantity confined by some $k_{max}$. That allows calculating the quantities, which are zero in
a vacuum state by definition, and non-zero in the ``exited''
states. The density of the created particles given by (\ref{mat})
is such a quantity. No dependence on $k_{max}$ appears for mean
values of such quantities because this consideration is, in fact,
equivalent to that using the Fock space and the normal operators
ordering.

   The second opportunity supported by cosmological arguments \cite{jcap} is to consider that $k_{max}$ exists really
 and is of the order of the Planck mass. That allows calculating the arbitrary quantities, which are nonzero in a vacuum state. The UV cut-off
 will figurate in the final result of these calculations \cite{jcap}.

\section{Informational content of a singularity}

Although the word ``information" is popular in different science
branches, the notion of information is not purely physical. The
most interrelated notion is the ``entropy," which is defined for
an assembly of the systems and equals zero for a pure quantum
state \cite{kak}. Saying about ``Everything,'' we imply that only
one single ``Everything'' exists, and it is a single universe
being in a single quantum state, i.e., its entropy equals zero.
Thus, other definition for the ``information content'' is
required.

A basic assumption could be accepted that a vacuum state contains
no information. The next point for introducing the information
could be the formula by Kullback-Leibler \cite{kullback} comparing
two probability distributions of a random variable in the
information theory. In fact, it is a measure in the functional
space of the functions $\bar \wp_k$.

One may use the mean value of the spectral energy density $\bar
\wp_k$ (\ref{stat})  of the created particles and, from the other
hand, use the vacuum value (see (\ref{mat}), (\ref{vac}) and
(\ref{def})) $\tilde \wp_k$ for the normalization. Thus, the
following prescription for the  information density can be taken:
\be
\mathcal I= \sum_{\bm k} \frac{\bar
\wp_k}{k}\ln\left(1+\frac{\bar\wp_k}{\tilde
\wp_k}\right)=\frac{1}{2\pi^2}\int_0^\infty  \frac{\bar
\wp_k}{k}\ln\left(1+\frac{\bar\wp_k}{\tilde \wp_k}\right)k^2dk.
\label{inf}
\ee
As can see, it is the distance in a functional space between the
function ${\bar \wp_k}$ and some singled out function ${\bar
\wp_k}=0$, whereas the  ${\tilde \wp_k}$ given by Eq.
(\ref{vac})does not correspond to the real particles, but rather
to virtual ones. A logarithmic scale is used in (\ref{inf}) like
the star luminosity in astrophysics or the decibels in acoustics,
and besides, the definition of the entropy in statistical physics.
\begin{figure}[th]
\includegraphics[width=7cm]{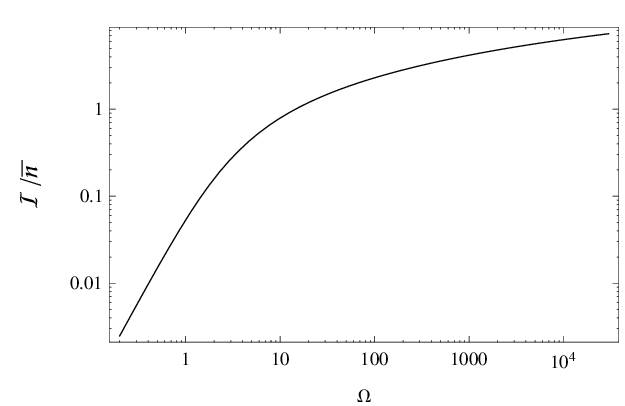}
\caption{\normalfont Amount of information per one particle,
depending on the parameter of the wave packet
(\ref{spec}).}\label{fig2}
\end{figure}
Having the density of created particles \be \bar n=\sum_{\bm k}
\frac{\bar \wp_k}{k},
\ee
the information  $\mathcal I/\bar n$ per one created particle can
be calculated. Substitution of the values (\ref{spec}) into
(\ref{stat}) results in
\be
a^4\bar \wp_k=\frac{k}{2} \left(\sqrt{1+ \Omega ^2 e^{-2 k
L}}-1\right).
\label{finwp}
\ee
Using (\ref{finwp}) and the asymptotic vacuum value of  $a^4\tilde
\wp_k\approx k/2$ (see Eq. (\ref{vac})) gives the particle density
and the information density in a final form
\bea
a^4\bar n=\frac{1}{4\pi^2}\int_0^\infty  \left(\sqrt{1+ \Omega ^2
e^{-2 k L}}-1\right)k^2dk,\\
a^4\mathcal I=\frac{1}{8\pi^2} \int_0^\infty  \left(\sqrt{1+
\Omega ^2 e^{-2 k L}}-1\right)\ln\left(1+ \Omega ^2 e^{-2 k
L}\right)k^2dk.
\eea

Roughly, the particle density is of the order of $\bar
n\sim\frac{\Omega}{a^4L^3}$, whereas the information density is
$\mathcal I\sim \frac{\Omega\ln\Omega}{a^4L^3}$. In the example
considered, an informational content per created particle
increases only logarithmically with $\Omega$ as shown in Fig.
\ref{fig2}. Although the particles' density could be increased by
decreasing $ L $, this does not increase information content per
particle.

Here we consider only some simple class of the quantum states and
a straightforward definition of the information. In principle,
other definitions of information are of interest. Let us propose
to Reader the information formula accounting for quantum states
explicitly:
\bea
\mathcal I=\int \left(C(\{P\})\ln\frac{C(\{P\})}{\tilde
C(\{P\})}\right)^*\sum_{\bm k} \frac{\hat \wp_{\bm
k}}{k}\,C(\{P\})\ln\frac{C(\{P\})}{\tilde C(\{P\})}\mathcal D
P_{\bm q}\mathcal D P_{\bm q}^*,
\label{inf1}
\eea
where $\tilde C(\{P\})$ is the wave packet (\ref{pak}) with
$\Delta_{\bm k}^\prime=k/\mathcal H^2$, $\Delta_{\bm
k}^{\prime\prime}=0$ producing no real particles in a future
asymptotically. On the one hand, the formula (\ref{inf1}) uses the
Kullback-Leibler idea but compares the quantum states with the
state $\tilde C(\{P\})$ giving no particles. On the other hand, it
contains features of the expression for the appeared particles'
mean density:
\be
\fl \bar n= \int \left(\left(C(\{P\})\right)^*\sum_{\bm k}
\frac{\hat \wp_{\bm k}}{k}\,C(\{P\})-\left(\tilde
C(\{P\})\right)^*\sum_{\bm k} \frac{\hat \wp_{\bm k}}{k}\,\tilde
C(\{P\})\right)\mathcal D P_{\bm q}\mathcal D P_{\bm q}^*.
\label{apppat}
\ee
The calculations, along with the definition (\ref{inf1}) seem more
complex, and we defer the examination of this issue to a later
date.

\section{Conclusions and Discussion}

To summarize, we considered an illustrative example within a
comparatively simple approach illustrating how the information
about ``Everything" could be stored at the singularity. We
demonstrated that the momentum wave packet could be defined at the
cosmological singularity so that: 1) some amount of matter can be
``placed" at the singularity, and, thereby, 2) some information
can be encoded into it. It is not creating the particles from
vacuum \cite{Par69,SexUrb69,ZelSta71,frol}, which is widely
considered at 1960th. This phenomenon produces a very low amount
of matter for the power-law of universe expansion, including the
linear expansion in cosmic time \cite{nl}. On the other hand, a
vacuum could be defined only after the moment when the field
oscillators begin to oscillate, which is relatively far from the
singularity. In contrast, in the approach considered, it seems
evident that one could place any amount of matter and information
into the singularity.

In the second section, the pure Heisenberg picture was considered
in which a Heisenberg operator evolves under some time-independent
state. This state has been constructed (or "enumerated") in terms
of momentum operators existing at the singularity. It allows
claiming that the state is set at the singularity and determines
the universe's matter and informational content.

The quasi-Heisenberg picture of the third section is more drastic
because the hyperplane $a=0$, figuring in the mean values'
definition, points that a state is set directly at the
singularity. It occurs that both formalisms are equivalent in the
framework of the simplified model defined on the classical
background. In the general quasi-Heisenberg picture of quantum
gravity, the quantum state would inevitably be set at the
singularity due to the hyperplane $a=0$ in the Klein-Gordon scalar
product.

In this work, the problem of the source of matter and information
is considered for the relatively narrow class of the theories with
singularity and without inflation. It is interesting to discuss a
broader class of theories. For instance, the loop-quantum gravity
(LQG) is based on the version of the Wheeler-DeWitt equation and
operates mainly in a timeless manner \cite{rov,lp}, i.e., the time
is recovered only quasi-classically, or not used at all by
considering the transition amplitudes. Besides, the ``bounces''
arise on the quasi-classical level, which excludes singularity. In
such a case, a state determining matter and information in the
universe is not connected with the singularity or with some moment
generally, and exists "elsewhere." Due to the combinatorial
features of LQG, it could be interesting to consider some
definitions of the information for such a state of the universe.
It should be noted that a similar problem of the entropy of a
black hole is was solved in LOQ \cite{lp}, considering a black
hole as an open system connected with the environment through the
horizon.

The discussion of inflation within the above context is also
interesting because the initial inhomogeneities determining the
universe's history originate from vacuum fluctuations. At the same
time, the enormous expansion during inflation hides the previous
universe's history. In some sense, inflation is analogous to
singularity because field oscillators do not oscillate near a
singularity, at the same time, when the scale of inhomogeneity
becomes greater horizon during inflation, the field oscillators
cease to oscillate also.    Then,  during the inflationary
universe expansion, the inhomogeneities cross the horizon (i.e.,
enter into the causality-connected region) and begin to oscillate
again. It is interesting to note, that in the Milne-like
cosmologies \cite{mil,John,loh1,conf0,levy,fm,lew,Singh,plasm} the
events always remain within the horizon.

\section*{References}

\bibliography{spec}

\end{document}